# SET-MEMBERSHIP LOCALIZATION
# VIA RANGE MEASUREMENTS

GIUSEPPE C. CALAFIORE*

**Abstract.** In this paper we discuss a classical geometrical problem of estimating an unknown point's location in $\mathbb{R}^n$ from several noisy measurements of the Euclidean distances from this point to a set of known reference points (anchors). We approach the problem via a set-membership methodology, in which we assume the distance measurements to be affected by unknown-but-bounded errors, and we characterize the set of all points that are consistent with the measurements and their assumed error model. This set is nonconvex, but we show in the paper that it is contained in a region given by the intersection of certain closed balls and a polytope, which we call the *localization set*. Then, we develop efficient methods, based on convex programming, for computing a tight outer-bounding set of simple structure (a box, or an ellipsoid) for the localization set, which then acts as a guaranteed set-valued location estimate. The center of the bounding set also serves as a point location estimate. Related problems of inner approximation of the localization set via balls and ellipsoids are also posed as convex programming problems. Different from existing methods based on semidefinite programming relaxations of a nonconvex cost minimization problem, our approach is direct, geometric and based on a polyhedral set of points that satisfy pairwise differences of the measurement equations.

**1. Introduction.** Accurate localization is a fundamental problem arising in various applications, including wireless sensor networks, robotics, and navigation systems. One widely used approach for localization is multilateration, which aims at determining the position of an object (e.g., vessel, aircraft, robot, etc.) based on range (distance) measurements from multiple known reference points, called beacons or anchors. Depending on the application, the distance measurements can be inferred from the time of arrival (TOA) or time difference of arrival (TDoA) of signals from beacons, from direct laser measurements, from received signal strength, etc.

There exists a vast literature on multilateration methods, with classical results dating back to the 1980s and early 1990s [2, 11]; we direct the reader to the recent review paper [27] for a complete survey of the topic of multilateration, its applications, and pointers to relevant references. In the typical approach to multilateration, an objective criterion is constructed which accounts for the sum of residuals between measurements and point-to-beacon distances, and such objective is then minimized with respect to the unknown position by means of optimization techniques. The main issue in this respect is that the objective takes typically the form of a nonlinear Least Squares problem, named *range-based LS*, or of a nonlinear Maximum Likelihood estimation problem, and this problem is nonconvex and difficult to solve in general. Certain approaches, see, e.g., [4], hence apply convex relaxation techniques to obtain an approximate problem that is amenable to solution via semidefinite programming solvers. A different approach, called the *squared-range-based least squares* (SR-LS), applies instead a Least Squares methodology to the *squared* distance measurements, see again [4]. This also results in a nonconvex problem formulation, which is amenable to approximate solution via relaxation. Along this line of minimizing a sum of measurement equation errors via convex relaxation, in the more general context of sensor network localization, are also the contributions in [7], where semidefinite programming (SDP) relaxations are proposed, in [23] where a second-order cone (SOC) relaxation is studied, and in [19] where a SOC approach is proposed, considering also uncertainty on the anchors' locations. A more classical approach is also to solve directly the nonconvex optimization problem by means of iterative methods such as the Gauss-Newton algorithm, see, e.g., [9, 31]. Also, it is well known that squared distance measurement equations can be transformed into a set of *linear* equations by taking pairwise differences between equations, which makes the quadratic terms elide, see, e.g., [11, 17]. The resulting set of linear equations can then be solved via linear Least Squares.

---

*Dipartimento di Elettronica e Telecomunicazioni, Politecnico di Torino, Italy. Email: giuseppe.calafiore@polito.it

These known localization methods typically rely on a stochastic description of measurement noise, assuming it follows a known statistical distribution. Further, they all focus on computing *one single* point estimate of the unknown location. In many practical scenarios, however, noise characteristics may be uncertain or difficult to model accurately, and this may lead to degraded localization performance. Further, it may be of practical interest not only to compute a point estimate, but actually to estimate an entire region in space in which the object of interest may be located, based on the measurements information and on the a-priori knowledge on the noise characteristics.

In this paper we explore a *set-membership* approach to localization, which offers an alternative framework that does not rely on probabilistic assumptions. Instead, it assumes that measurement errors lie within known intervals, allowing for a set-based analysis of the localization problem. Error intervals are indeed more easily and naturally obtainable in practice than a full stochastic description of the error distribution, see Remark 1 below for further discussion. By leveraging set theory and constraint-based formulations, the set-membership approach provides estimates of the *region* where the target object is guaranteed to reside, rather than merely a point estimate. Such guarantees make set-membership localization particularly attractive for safety-critical applications requiring robust and reliable position estimation.

The literature on set-membership approaches to localization via multilateration seems however to be rather scarce. For instance, the approach in [29] is limited to two-dimensional localization, and proposes distributed set-membership filtering based on a received signal strength indicator (RSSI) to dynamically reduce the localization error. In [26] the authors propose a set-membership approach to improve the positioning accuracy and stability of a Visible Light Positioning system in two dimensions, which is solved by means of a computationally-intensive interval analysis-based set inversion algorithm. Closer to our approach, in [21] the authors consider interval errors in the measurement equations and then seek for the positions estimate which minimizes the worst-case estimation error over all feasible positions, i.e., over all positions that are compatible with the error bounds. The resulting problem is a nonconvex optimization problem, which is solved by means of semidefinite relaxation techniques. The approach of [21], however, while considering interval errors and worst-case errors, is still aimed at computing one single point estimate, and it is based on nonconvex problem formulations that are dealt with by means of relaxations.

In the present work, we first observe that under an interval error setting, plain distance measurements equations and squared distance measurements equations are equivalent and lead to a unified model of the form (2.3), which also accommodates both absolute (additive) and relative (multiplicative) error terms. Based on this general model, the $m$ nonlinear equations related to the distance or squared distance measurements from a point $x$ to the $m$ beacons are transformed into a set of $p = m(m-$ 1)/2 linear equations obtained by taking all pair-wise differences among the original equations. These linear equations are subject to polytopic (and non independent) errors. In Section 3 we characterize explicitly a convex localization set $\mathcal{X}$ that is guaranteed to contain the original set $\mathcal{X}_{\text{true}}$ of all point consistent with the original nonlinear measurement equations. Then, in Section 4 we show how to compute a maximal ellipsoid inscribed in $\mathcal{X}$; the center of such ellipsoid can be taken as a central estimate of the unknown target's position. Notably, such inscribed ellipsoid can be computed efficiently via convex programming; in particular, we need to solve a convex second-order cone program (SOCP) for finding an inscribed ball (which is a particular case of an ellipsoid), or a semidefinite program (SDP) for finding a general inscribed ellipsoid. The key Section 5 discusses instead how we can compute minimal *outer* bounding sets for $\mathcal{X}$, and hence for $\mathcal{X}_{\text{true}} \subseteq \mathcal{X}$. We discuss in particular a simple technique for computing a bounding box $H_{\text{outer}} \supseteq \mathcal{X}$ via SOCP, and a technique for computing a bounding ellipsoid via SDP. Further aspects are then investigated in Section 6. In particular, Section 6.1 discusses a local approximation approach based on convex quadratic programming for seeking a point in $\mathcal{X}_{\text{true}}$, while Section 6.2 discusses how we can deal with the issue of outliers or error realizations that do not respect the assumed bounds. Although many practical localization problems are posed in two-dimensional or three-dimensional spaces, we remark that the methodology proposed in the present paper works in spaces of any finite dimension. The proposed approach is tested numerically on random scenarios in Section 7. Conclusions are drawn in Section 8.

**1.1. Further discussion on related work.** Some recent studies have explored modeling frameworks that are somewhat related to, but distinct from, the proposed approach. For example, Yu et al. [30] introduced Ellipsoid SLAM, where location uncertainty is modeled via bounding ellipsoids in the SLAM context, demonstrating the inadequacy of Gaussian noise assumptions in many real-world scenarios and the effectiveness of an unknown-but-bounded (UBB) approach. The SLAM problem discussed in [30], however, is substantially different from the one we discuss in the present paper. Indeed, SLAM is based on recursive linearization, and hence approximation, of the state and measurement equations. Linearization is a local approach which introduces non-convex residual errors that are not captured by the ellipsoidal bound. Therefore, the propagated ellipsoid bears no theoretical guarantee of actually containing the system state: it is merely an approximation, often under-approximating the true feasible region. This is distinctively different from our approach, which provides guaranteed overbounding sets. Similarly, in [18] the authors propose a method for multiple-source localization based on acoustic energy measurements. In this case, the measurement equation depends on the distance from acoustic sources according to the sum of inverse power laws, and the measurement error is assumed to be bounded in a given interval. A bounding ellipsoid is computed for each source based on linearization of the measurement equation and conservative on-line bounding of the linearization residual.

Zhang et al. [32] proposed a set-membership approach to state estimation using zonotopes (a special type of polytopes) rather than ellipsoids. While [32] makes again a case for the use of UBB noise models in a state estimation setting, its approach only applies to *linear* dynamical systems, and thus can not deal with nonlinear measurement models such as the one arising in the multilateration context we discuss in this paper. Wang et al. [24] deal with ellipsoidal state estimation in a nonlinear setting, but instead of classical linearization they propose the reformulation of the state-bounding problem as a semi-infinite optimization problem which is then solved approximately via constraint sampling, thus again lacking formal guarantees of state containment.

The above-mentioned approaches deal with dynamic state estimation and are based on recursive local approximation (linearization). They all require an "initial guess" localization set to be provided, e.g., in the form of an ellipsoid, so to allow the linearization procedure to start. The problem we consider in this paper tackles instead the localization problem from a global point of view, with no initial guess available for performing a linearization approach. In a sense, the method discussed in the present paper can be seen as a technique for determining the initial localization set needed by recursive, linearization-based, methods.

*Notation.* We use a standard notation for vectors and matrices: $x \in \mathbb{R}^n$ denotes a column vector of dimension $n$, and $A \in \mathbb{R}^{m,n}$ denotes a matrix with $m$ rows and $n$ columns. $A^\top$ denotes the transpose of $A$ while $A^\dagger$ denotes the Moore-Penrose pseudo-inverse of $A$. We denote by $\mathbb{S}^n$ the space of $n \times n$ real symmetric matrices. We denote by $I_n$, $\mathbf{1}_n$ the $n \times n$ identity matrix and a vector of all ones of dimension $n$, respectively; the suffix specifying the dimension can be omitted when easily inferable from context. $\mathcal{N}(A)$ denotes the null space (or kernel) of $A$, $\mathcal{R}(A)$ denotes the range space of $A$, $\mathcal{N}^\perp(A)$ denotes the orthogonal complement of $\mathcal{N}(A)$. When involving vectors, inequalities are intended element-wise. The notation $A \succeq 0$ means that $A$ is symmetric and positive semidefinite.

**2. Problem setup.** We are given $m$ anchor points with known coordinates in a given reference frame, described by vectors $a^{(i)} \in \mathbb{R}^n$, $i = 1, \ldots, m$, and we collect $m$ measurements related to the Euclidean distance from point $x \in \mathbb{R}^n$ to each anchor point. Two types of measurements are typically considered, one based on the squared distance and one based on the distance itself:

- Squared distance measurements:

$$(2.1) \qquad y_i = \|x - a^{(i)}\|_2^2 + \epsilon_i, \quad i = 1, \ldots, m.$$

- Plain distance measurements:

$$(2.2) \qquad z_i = \|x - a^{(i)}\|_2 + e_i, \quad i = 1, \ldots, m.$$

Each of these models can be justified based on the specific application; e.g., time-of-flight measurements that relate distance to the travel time of a signal from the anchor to the point typically return plain distance measurements, while power attenuation measurements typically return squared distance measurements, see, e.g., [27].

The error terms in the above range measurement models are here considered to be unknown-but-bounded, rather than stochastic, see Remark 1 for further motivation of this assumption. In particular, we accommodate in our model both *absolute* interval errors where $\epsilon_i \in [\epsilon_i^-, \epsilon_i^+]$, $i = 1, \ldots, m$, in model (2.1) or $e_i \in [e_i^-, e_i^+]$, $i = 1, \ldots, m$, in model (2.2), and *relative* interval errors where $\epsilon_i = \|x - a^{(i)}\|_2^2 \tilde{\epsilon}_i$, $\tilde{\epsilon}_i \in [\tilde{\epsilon}_i^-, \tilde{\epsilon}_i^+]$, $i = 1, \ldots, m$, in model (2.1) or $e_i = \|x - a^{(i)}\|_2 \tilde{e}_i$, $\tilde{e}_i \in [\tilde{e}_i^-, \tilde{e}_i^+]$, $i = 1, \ldots, m$, in model (2.2). We let $\epsilon$, $\epsilon^-$, $\epsilon^+$ denote the vectors whose entries are $\epsilon_i$, $\epsilon_i^-$, $\epsilon_i^+$, respectively, for $i = 1, \ldots, m$, and we define similarly vectors $\tilde{\epsilon}, \tilde{\epsilon}^-, \tilde{\epsilon}^+, e, e^+, e^-, \tilde{e}, \tilde{e}^+, \tilde{e}^-$. We assume that $\tilde{\epsilon}_i^- > -1$, $\tilde{e}_i^- > -1$ for all $i$, and that $y_i \geq \epsilon_i^+$, $z_i \geq e_i^+$ for all $i$.

We next observe that both model (2.1) and model (2.2), with both absolute or relative errors, all reduce to a single equivalent model of the form

$$(2.3) \qquad \|x - a^{(i)}\|_2^2 = \xi_i, \quad \xi_i \in [\xi_i^-, \xi_i^+], i = 1, \ldots, m,$$

where the upper and lower bounds on $\xi_i$ differ depending on the specific model and type of error (absolute or relative). In particular, it can be easily verified that:

- For model (2.1) with absolute errors, it holds that

$$(2.4) \qquad \xi_i = y_i - \epsilon_i, \quad \xi_i^- \doteq y_i - \epsilon_i^+, \, \xi_i^+ \doteq y_i - \epsilon_i^-, \quad i = 1, \ldots, m.$$

- For model (2.1) with relative errors, it holds that

$$(2.5) \qquad \xi_i = \frac{y_i}{1+\tilde{\epsilon}_i}, \quad \xi_i^- \doteq \frac{y_i}{1+\tilde{\epsilon}_i^+}, \, \xi_i^+ \doteq \frac{y_i}{1+\tilde{\epsilon}_i^-}, \quad i = 1, \ldots, m.$$

- For model (2.2) with absolute errors, it holds that

$$(2.6) \quad \xi_i = (z_i - e_i)^2, \quad \xi_i^- \doteq (z_i - e_i^+)^2, \, \xi_i^+ \doteq (z_i - e_i^-)^2, \quad i = 1, \ldots, m.$$

- For model (2.2) with relative errors, it holds that

$$(2.7) \quad \xi_i = \frac{z_i^2}{(1+\tilde{e}_i)^2}, \quad \xi_i^- \doteq \frac{z_i^2}{(1+\tilde{e}_i^+)^2}, \, \xi_i^+ \doteq \frac{z_i^2}{(1+\tilde{e}_i^-)^2}, \quad i = 1, \ldots, m.$$

We let $\xi$, $\xi^-$, $\xi^+$ denote the vectors containing entries $\xi_i$, $\xi_i^-, \xi_i^+$, respectively, for $i = 1, \ldots, m$, and we let $H$ denote the hyperrectangle

$$(2.8) \qquad\qquad H = \{\xi \in \mathbb{R}^m : \xi^- \leq \xi \leq \xi^+\}.$$

In the sequel we focus only on the general model (2.3), which is representative of all four measurement models discussed above. According to (2.3), each measurement confines the possible locations $x$ in a hyperspherical shell region

$$(2.9) \qquad\qquad \mathcal{S}_i = \{x \in \mathbb{R}^n : \sqrt{\xi_i^-} \leq \|x - a^{(i)}\|_2 \leq \sqrt{\xi_i^+}\}.$$

The set $\mathcal{X}_{\text{true}}$, containing all $x$ locations that are compatible with all the uncertain measurements, is the intersection of the $\mathcal{S}_i$ regions:

$$\mathcal{X}_{\text{true}} = \bigcap_{i=1,\ldots,m} \mathcal{S}_i.$$

$\mathcal{X}_{\text{true}}$ is in general a nonconvex set. Figure 1 shows an example in $\mathbb{R}^2$ with three anchors, where each $\mathcal{S}_i$ is a circular annulus.

REMARK 1 (Interval noise model). An unknown-but-bounded assumption on measurement errors is appropriate, for instance, when one has physical sensor error limits, or when the measurement device provides a certified accuracy or resolution, e.g., a LiDAR range sensor with $\pm 3$cm repeatability, see, e.g., [22, 15]. Error bounds can be rigorously derived from the physics of the underlying sensor mechanisms or from manufacturer calibration certificates, see, e.g., [14]. Also, in situations in which one can not confidently identify a noise distribution (non-stationary sensors, aging equipment, variable environments), a deterministic bound is more credible than assuming, say, Gaussianity of the measurement errors. This is typical in early-stage system identification or in field robotics where environmental conditions fluctuate. In practice, bounds on errors can be obtained from calibration data, manufacturer specifications, or from truncated statistical models (e.g., $3\sigma$ Gaussian bounds).

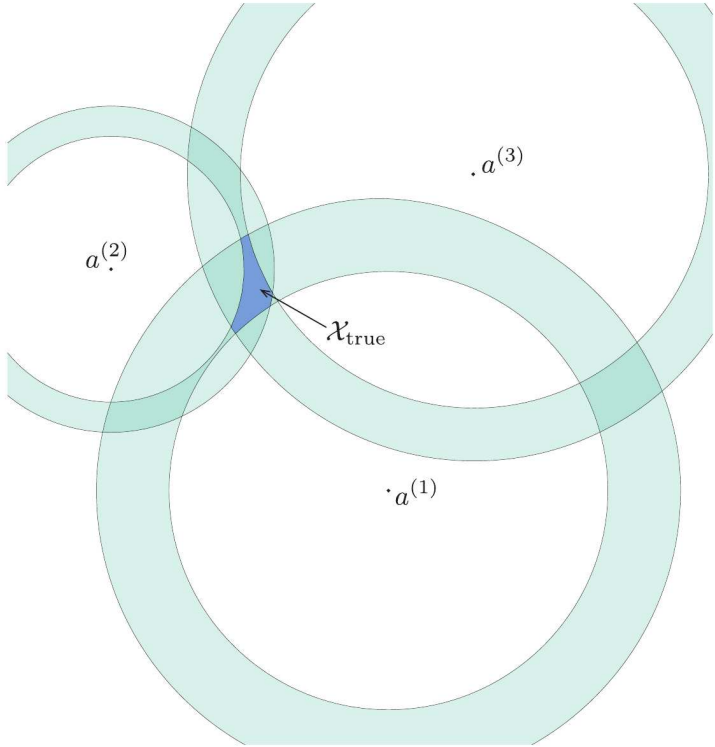

Fig. 1. *Localization with three anchors in $\mathbb{R}^2$.*

In safety-critical systems (aerospace, autonomous driving, process control) one needs guarantees for any admissible noise realization. In such cases, the UBB model allows one to prove that constraints on states or outputs hold for all noise realizations within the bounds, a statement which is not obtainable via purely stochastic models. In our context, the use of UBB noise and a set-membership approach leads to deriving an outer bounding set which is *guaranteed* to contain our localization target.

We also remark that the specific interval UBB model used in eqs. (2.1), (2.2) appears more suitable in our context that an ellipsoidal UBB error model. Indeed, it seems natural to assume that measurement errors in eq. (2.1) or in eq. (2.2) are independent from one another, since they depend only on the characteristics of the distance measurement device, and this is reflected well in the assumptions that they lie in independent intervals. An ellipsoidal model for the measurement errors would instead imply a mutual dependence among the measurement errors (e.g., the amplitude of errors $\epsilon_1, \ldots, \epsilon_{m-1}$ constrains the amplitude of error $\epsilon_m$), which would be unnatural in our context. ⋆

In the next section we introduce a convex region $\mathcal{X}$, called the *localization set*, that provides a certified outer bound for the true feasible set $\mathcal{X}_{\text{true}}$ of locations compatible with the uncertain measurements (note that $\mathcal{X}_{\text{true}}$ is in general nonconvex). Specifically, $\mathcal{X}$ is defined as the intersection of the $m$ measurement closed balls and the difference-of-measurements (d.o.m.) polyhedron $\mathcal{X}_d$.

## 3. The d.o.m. polyhedron $\mathcal{X}_d$ and localization set $\mathcal{X}$.

### 3.1. The difference of measurements graph.
Observe that the $i$th measurement equation (2.3) can be expanded to

$$\|x\|_2^2 - 2x^\top a^{(i)} + \|a^{(i)}\|_2^2 = \xi_i$$

Then, subtracting the $j$th measurement equation from the $i$th, for $i \neq j$, we obtain that

$$\begin{aligned}
&\|x\|_2^2 - 2x^\top a^{(i)} + \|a^{(i)}\|_2^2 - \|x\|_2^2 + 2x^\top a^{(j)} - \|a^{(j)}\|_2^2 \\
&= -2x^\top(a^{(i)} - a^{(j)}) + \|a^{(i)}\|_2^2 - \|a^{(j)}\|_2^2 \\
&= \xi_i - \xi_j,
\end{aligned}$$

where we observe that the quadratic terms in $x$ disappear. Considering then all pairs of measurement equations, with $i < j$, we obtain $p = m(m-1)/2$ linear equations in $x$ of the form

$$(3.1) \qquad 2x^\top(a^{(i)} - a^{(j)}) = \|a^{(i)}\|_2^2 - \|a^{(j)}\|_2^2 - (\xi_i - \xi_j),$$

for all $i, j = 1, \ldots, m$, $i < j$. We assume without loss of generality that the $p$ pairs are ordered as $(1,2), (1,3), \ldots, (1,m), (2,3), \ldots, (2,m), \ldots, (m-1, m)$. We next define

$$(3.2) \qquad \nu = E\xi,$$

where $E \in \mathbb{R}^{p,m}$ is a matrix whose $k$th row has the form $(e^{(i)} - e^{(j)})^\top$, where $e^{(i)} \in \mathbb{R}^m$ is the $i$th standard basis vector (i.e., composed of all zeros, except for a one in position $i$), and $(i,j)$, $i < j$, form the $k$th pair of anchors.

REMARK 2. There is a natural graph-theoretic interpretation of our problem structure. The $m$ anchors form the vertices of a graph, and we say there is an edge $(i, j)$ from node $j$ to $i$ whenever we consider the corresponding difference of measurements eq. (3.1). Considering all difference of measurements $(i, j)$ for all $j > i$ thus forms a connected graph $\mathcal{G}$ with $m$ vertices and $p = m(m-1)/2$ edges. We observe that, according to its definition, $E^\top \in \mathbb{R}^{m,p}$ represents the incidence matrix of such graph. ⋆

Letting

$$\begin{aligned}
A &\doteq [a^{(1)} \cdots a^{(m)}] \in \mathbb{R}^{n,m}, \\
\omega &\doteq [\|a^{(1)}\|_2^2 \cdots \|a^{(m)}\|_2^2]^\top \in \mathbb{R}^m,
\end{aligned}$$

one can verify that eqs. (3.1) can be rewritten compactly as

$$(3.3) \qquad 2EA^\top x = E\omega - \nu, \quad \nu = E\xi$$

where

$$EA^\top = \begin{bmatrix} (a^{(1)} - a^{(2)})^\top \\ \vdots \\ (a^{(m-1)} - a^{(m)})^\top \end{bmatrix}, \quad E\omega = \begin{bmatrix} \|a^{(1)}\|_2^2 - \|a^{(2)}\|_2^2 \\ \vdots \\ \|a^{(m-1)}\|_2^2 - \|a^{(m)}\|_2^2 \end{bmatrix}.$$

**3.2. The difference-of-measurements polyhedron.** We next characterize a convex polyhedral set[1] that is guaranteed to contain $\mathcal{X}_{\text{true}}$.

PROPOSITION 3.1. *Let $V \doteq \{\nu \in \mathbb{R}^p : \nu = E\xi, \xi \in H\}$. The set*

$$\mathcal{X}_d \doteq \{x \in \mathbb{R}^n : 2EA^\top x = E\omega - \nu,\ \nu \in V\}$$

*is a polyhedron, and it holds that $\mathcal{X}_{\text{true}} \subseteq \mathcal{X}_d$.*

**Proof.** We first observe that for any $x \in \mathcal{X}_{\text{true}}$, by definition, all eqs. (2.3) hold for some $\xi \in H$, and hence the relative difference equations (3.1), which are derived from (2.3), also hold, which means that $x \in \mathcal{X}_d$, and therefore $\mathcal{X}_{\text{true}} \subseteq \mathcal{X}_d$. Next, observe that $V$ is a polytope, since it is the image of the hyperectangle $H$ through a linear map, see, e.g., [13]. Similarly, $\mathcal{X}_d$ is a polyhedron since it is the preimage of a polytope through an affine map, see, e.g., ex. 3.9 in [6]. □

It follows from Proposition 3.1 that the set $\mathcal{X}_d$ is characterized via a set of linear equalities and inequalities as $\mathcal{X}_d = \{x \in \mathbb{R}^n : 2EA^\top x = E\omega - E\xi, \xi \in H\}$. Finding a point in $\mathcal{X}_d$, or determining that no such point exists, amounts to solving a linear programming (LP) feasibility problem. Observe that while $\xi$ belongs to an orthotope $H$, that is, the elements of $\xi$ are bounded in independent intervals, this is not true for $\nu = E\xi$, whose entries are instead dependent. We next give an alternate description of $V$ and hence of $\mathcal{X}_d$.

PROPOSITION 3.2. *Let $\tilde{V} \doteq \{\nu \in \mathbb{R}^p : \nu \in \mathcal{R}(E), E^\dagger \nu + \alpha \mathbf{1} \in H, \alpha \in \mathbb{R}\}$. Then, $\tilde{V} = V$.*

**Proof.** As observed in Remark 2 matrix $E \in \mathbb{R}^{p,m}$ represents the (transpose) incidence matrix of the difference-of-measurements graph $\mathcal{G}$. Since this graph is connected by construction, it holds that $E$ has rank $m-1$, with nullspace $\mathcal{N}(E) = \{\alpha \mathbf{1}, \alpha \in \mathbb{R}\}$; see, e.g., Lemma 2.2 in [3]. Now, suppose first that $\nu \in V$. Then $\nu = E\xi$ for some $\xi$, which implies $\nu \in \mathcal{R}(E)$, and all $\xi$ compatible with this equation are parametrized by $\xi = E^\dagger \nu + \alpha \mathbf{1}$ for $\alpha \in \mathbb{R}$, therefore it must be $E^\dagger \nu + \alpha \mathbf{1} \in H$ for some $\alpha$, which implies that $\nu \in \tilde{V}$. Conversely, suppose $\nu \in \tilde{V}$. Then, $\nu \in \mathcal{R}(E)$ and we define $\xi = E^\dagger \nu + \alpha \mathbf{1} \in H$, for which it holds that

$$E\xi = EE^\dagger \nu + \alpha E \mathbf{1} = \nu,$$

since $\mathbf{1}$ is in the nullspace of $E$ and $EE^\dagger \nu = \nu$ due to the fact that $EE^\dagger$ is the orthogonal projector onto $\mathcal{R}(E)$, but $\nu \in \mathcal{R}(E)$ hence the projection coincides with $\nu$; so we showed that $\nu \in V$. This concludes the proof that $V$ and $\tilde{V}$ coincide. □

PROPOSITION 3.3. *The set $\mathcal{X}_d$ is given by*

(3.4) $$\mathcal{X}_d = \{x \in \mathbb{R}^n : \xi^- \leq Q\omega - 2QA^\top x + \alpha \mathbf{1} \leq \xi^+, \alpha \in \mathbb{R}\}$$

*where*

$$Q \doteq E^\dagger E = \frac{1}{m}\bigl(mI_m - \mathbf{1}_m \mathbf{1}_m^\top\bigr).$$

**Proof.** From the definition, $\mathcal{X}_d = \{x \in \mathbb{R}^n : (E\omega - 2EA^\top x) \in V\}$. By applying Proposition 3.2 we obtain the equivalent representation

$$\mathcal{X}_d = \{x \in \mathbb{R}^n : E\omega - 2EA^\top x \in \mathcal{R}(E),\ E^\dagger(E\omega - 2EA^\top x) + \alpha \mathbf{1} \in H, \alpha \in \mathbb{R}\}.$$

---
[1]A convex polyhedral set, or polyhedron, is the intersection of a finite number of closed half-spaces.

However, the first of these conditions, i.e., $E\omega - 2EA^\top x \in \mathcal{R}(E)$, is always satisfied, therefore we have that $\mathcal{X}_d = \{x \in \mathbb{R}^n : E^\dagger(E\omega - 2EA^\top x) + \alpha\mathbf{1} \in H, \alpha \in \mathbb{R}\}$, that is $\mathcal{X}_d = \{x \in \mathbb{R}^n : Q\omega - 2QA^\top x + \alpha\mathbf{1} \in H, \alpha \in \mathbb{R}\}$, where $Q \doteq E^\dagger E \in \mathbb{R}^{m,m}$ is the unique orthogonal projector matrix (symmetric, idempotent) onto the range of $E^\top$, which coincides with the orthogonal complement of the kernel of $E$. Since $\mathcal{N}(E) = \{\alpha\mathbf{1}\}$, it can be verified that we have explicitly $Q = \frac{1}{m}\left(mI_m - \mathbf{1}_m\mathbf{1}_m^\top\right)$, which concludes the proof. $\square$

REMARK 3 (Emptiness, boundedness). Observe that elements $x \in \mathcal{X}_d$ are those who solve the system of linear equations

$$2EA^\top x = E\omega - E\xi, \quad \text{for some } \xi \in H.$$

Since the left-hand side of the above equation is in $\mathcal{R}(EA^\top)$ while the right-hand side is in $\mathcal{R}(E)$, and it holds in general that $\mathcal{R}(EA^\top) \subseteq \mathcal{R}(E)$, we have that $\mathcal{X}_d$ is *nonempty* if and only if

(3.5) $$\exists \xi \in H : E(\omega - \xi) \in \mathcal{R}(EA^\top),$$

a condition that can be checked via linear programming, since it amounts to finding $\xi \in H$ such that $U^\top E\xi = U^\top E\omega$, where $U$ contains by columns a basis for the nullspace $\mathcal{N}(AE^\top)$, being $\mathcal{R}(EA^\top) = \mathcal{N}^\perp(AE^\top)$. Notice that if the measurements are indeed generated by the model equation (2.1) or (2.2) for some true location $x_{\text{true}}$, and if the noise terms respect their bounds, then there surely exist at least one point in $\mathcal{X}_{\text{true}}$ (e.g., $x_{\text{true}}$ itself), and hence $\mathcal{X}_{\text{true}}$ is nonempty, thus $\mathcal{X}_d \supseteq \mathcal{X}_{\text{true}}$ is also nonempty. Emptiness of $\mathcal{X}_d$ can hence only happen if some measurement does not respect the assumed model equation, or if the error bounds are not respected, which may happen in practice in the presence of outliers or unexpected errors or failures in the measurement equipment; see Section 6.2 for a discussion on how to tackle this situation.

Set $\mathcal{X}_d$ is *unbounded* if (3.5) holds and $\mathcal{N}(EA^\top)$ is nontrivial, since obviously in such case $x$ may grow unbounded along any direction in $\mathcal{N}(EA^\top)$. Thus, the condition that $EA^\top$ has full rank guarantees boundedness of $\mathcal{X}_d$ (and hence of $\mathcal{X}_{\text{true}}$). Notice in particular that in the special but common case when $m = n+1$ and rank$(EA^\top) = n$ then $\mathcal{R}(EA^\top) = \mathcal{R}(E)$, hence $\mathcal{X}_d$ is surely bounded and nonempty. $\star$

REMARK 4 (Standard conditions). The typical situations we are interested in are those in which $n \geq 2$ and $m \geq n+1$, that is the number of anchors is larger than the space dimension. This assumption insures that $p = m(m-1)/2 \geq m > n$ so that matrices $A \in \mathbb{R}^{p,n}$ and $E \in \mathbb{R}^{p,m}$ are "tall" matrices. We also typically consider anchors that are in general position, so that $EA^\top$ is full column rank, i.e., rank$(EA^\top) = n$, which guarantees boundedness of $\mathcal{X}_d$, as mentioned in the previous remark. $\star$

**3.3. The localization set $\mathcal{X}$.** In Section 3.2 we characterized a polyhedron $\mathcal{X}_d$ which is guaranteed to contain $\mathcal{X}_{\text{true}}$. We next observe that, by definition, $\mathcal{X}_{\text{true}}$ is also contained in a convex region $\mathcal{H}$ given by intersection of $m$ closed balls

(3.6) $$\mathcal{H}_i \doteq \{x \in \mathbb{R}^n : \|x - a^{(i)}\|_2 \leq \sqrt{\xi_i^+}\},\ i = 1,\ldots,m, \quad \mathcal{H} \doteq \bigcap_{i=1}^m \mathcal{H}_i.$$

Defining

(3.7) $$\mathcal{X} \doteq \mathcal{X}_d \cap \mathcal{H}$$

we have that $\mathcal{X}_{\text{true}} \subseteq \mathcal{X}$, and we hence call $\mathcal{X}$ the *localization set*. This set provides a convex outer bounding of the true (and nonconvex) set $\mathcal{X}_{\text{true}}$ of possible solutions of the noisy measurement equations (2.3), and refines the outer bound set given by $\mathcal{X}_d$. The tightness of the bounding $\mathcal{X}_{\text{true}} \subseteq \mathcal{X}$ depends on anchors' configuration, target position and noise bounds. We do not provide formal results on the bounding tightness in this paper. However, numerical experiments shows that $\mathcal{X}$ typically becomes nearly coincident with $\mathcal{X}_{\text{true}}$ when anchors are well distributed around the target, while near colinear or coplanar or clustered anchors increase conservativeness, see an exemplification of this fact in Figure 2 and Figure 3. The ratio $\text{vol}(\mathcal{X})/\text{vol}(\mathcal{X}_{\text{true}})$ provides a useful measure of this conservativeness and it is estimated in the numerical experiments of Section 7.

Figure 2 shows an example in $\mathbb{R}^2$ with $m = 3$ anchors, where the gray polygon depicts $\mathcal{X}_d$ and the area contoured in green depicts $\mathcal{X}_{\text{true}}$. The boundaries of the balls $\mathcal{H}_i$ (i.e., the corresponding spheres) are depicted in blue. The localization set $\mathcal{X} = \mathcal{X}_d \cap \mathcal{H}$ is not shown directly in the picture, but it can be easily inferred visually by intersecting $\mathcal{X}_d$ with the discs $\mathcal{H}_i$, $i = 1, 2, 3$. Figure 3 shows the same target and relative bounds used in Figure 2, but with one anchor progressively displaced so to worsen the conditioning of the $EA^\top$ matrix (anchor points tend to be "nearer to colinearity"). We observe that the gray polygon depicting $\mathcal{X}_d$ becomes larger as the anchors' geometry worsens. If the anchors become exactly colinear, $\mathcal{X}_d$ would become unbounded, as discussed in Remark 3. Boundedness of the localization set $\mathcal{X} = \mathcal{X}_d \cap \mathcal{H}$ is however maintained in all cases, due to the intersection of $\mathcal{X}_d$ with the bounded balls $\mathcal{H}_i$, $i = 1, \ldots, m$.

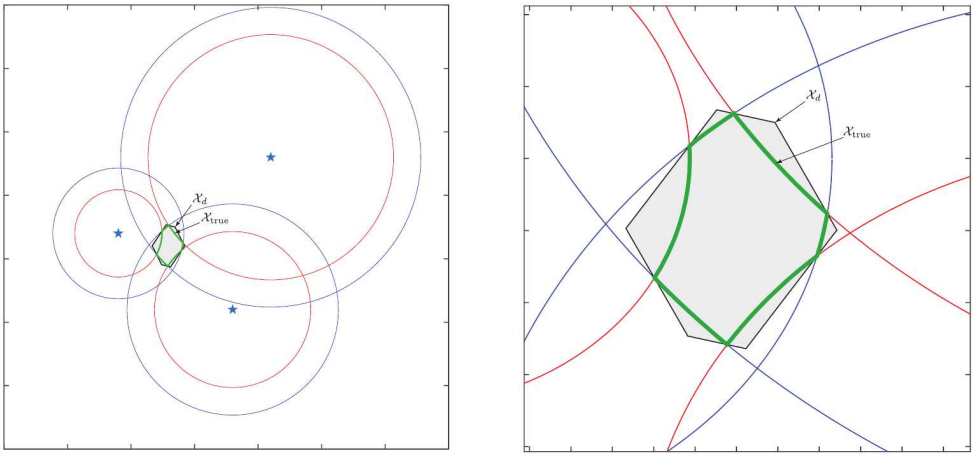

FIG. 2. *Sets $\mathcal{X}_{\text{true}}$ (bordered in green) and $\mathcal{X}_d$ (bordered in black and filled in gray) on a 2-dimensional example with three anchors. Left panel: full view with annular regions depicted in blue and red; Right panel: detail of the true non-convex localization set $\mathcal{X}_{\text{true}}$ and its bounding polytope $\mathcal{X}_d$.*

**4. Inner approximation of the localization set.** We next discuss how to compute simple sets included in $\mathcal{X}$ (the localization set), and hence obtain convenient central estimates of the location. Since $\mathcal{X}$ is a convex outer bound of the true feasible set $\mathcal{X}_{\text{true}}$, these inner approximations are guaranteed to lie in $\mathcal{X}$, but they are not necessarily subsets of $\mathcal{X}_{\text{true}}$. Section 5 discusses instead the main problem of finding

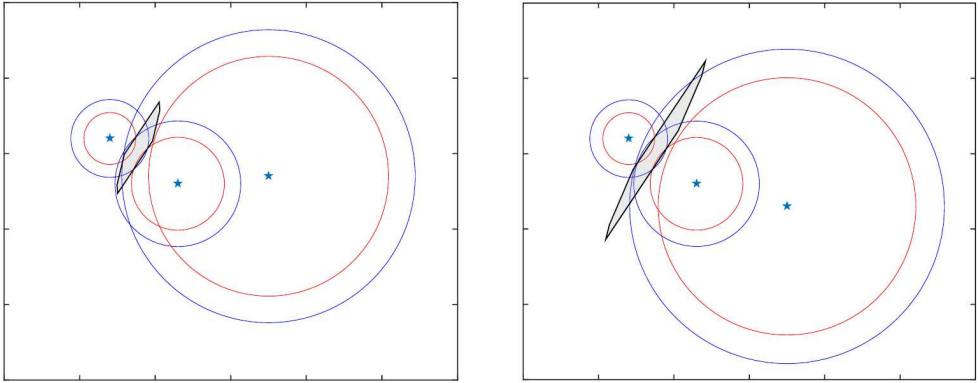

Fig. 3. *Modification of the polytope $\mathcal{X}_d$ with respect to Figure 2, as the right-most anchor moves so to worsen the conditioning of the anchors' geometry.*

an outer bounding set guaranteed to include $\mathcal{X}$, and hence $\mathcal{X}_{\text{true}}$.

**4.1. Computing an inscribed ball.** We take a Chebychev-type approach and discuss how to compute a large ball contained in $\mathcal{X}$. The central estimate will then be given by the center of such ball. Denoting the ball by

$$\mathcal{B}(c,r) = \{\xi \in \mathbb{R}^n : \xi = c + ru,\ c \in \mathbb{R}^n, \|u\|_2 \leq 1\},$$

where $c$ is the center and $r$ is the radius, we have that $\mathcal{B}(c,r) \subseteq \mathcal{X} = \mathcal{X}_d \cap \mathcal{H}$ if and only if $\mathcal{B}(c,r) \subseteq \mathcal{X}_d$ and $\mathcal{B}(c,r) \subseteq \mathcal{H}$. We next examine these two inclusions separately.

First, using the characterization in Proposition 3.3, we have that $\mathcal{B}(c,r) \subseteq \mathcal{X}_d$ if for any $x \in \mathcal{B}(c,r)$ there exist $\alpha \in \mathbb{R}$ such that

$$(4.1) \qquad \xi_i^- \leq -2[QA^\top]_i x + [Q\omega]_i + \alpha \leq \xi_i^+,\ i=1,\ldots,m,$$

where $[QA^\top]_i$ denotes the $i$th row of matrix $QA^\top$ and $[Q\omega]_i$ denotes the $i$th element of vector $Q\omega$. The problem here is that the $\alpha$ value may depend on $x$ (i.e., on $u$), hence if we impose the above conditions to hold for a single fixed and common $\alpha$ we obtain a conservative estimate. In such case we have that the above conditions hold if there exist $\alpha$ such that for $i=1,\ldots,m$,

$$\max_{\|u\|_2 \leq 1} -2[QA^\top]_i c - 2r[QA^\top]_i u + [Q\omega]_i + \alpha \leq \xi_i^+$$
$$\min_{\|u\|_2 \leq 1} -2[QA^\top]_i c - 2r[QA^\top]_i u + [Q\omega]_i + \alpha \geq \xi_i^-$$

Since $\max_{\|u\|_2 \leq 1} w^\top u = \|w\|_2$ and $\min_{\|u\|_2 \leq 1} w^\top u = -\|w\|_2$, the above conditions become explicitly

$$-2[QA^\top]_i c + 2r\|[QA^\top]_i\|_2 + [Q\omega]_i + \alpha \leq \xi_i^+$$
$$-2[QA^\top]_i c - 2r\|[QA^\top]_i\|_2 + [Q\omega]_i + \alpha \geq \xi_i^-.$$

We hence have that $\mathcal{B}(c,r) \subseteq \mathcal{X}_d$ if

$$-2QA^\top c + 2rz + Q\omega + \alpha \mathbf{1} \leq \xi^+$$
$$-2QA^\top c - 2rz + Q\omega + \alpha \mathbf{1} \geq \xi^-,$$

where $z \in \mathbb{R}^m$ is such that $z_i = \|[QA^\top]_i\|_2$, $i = 1, \ldots, m$. As discussed above, however, these conditions are conservative due to the fact that we considered a fixed $\alpha$ for all points in the ball. The situation can be improved (i.e., a *larger* inscribed ball can be obtained) by allowing $\alpha$ to depend on $x = c + ru$, hence on $u$. Specifically, we consider an affine dependence of the form

$$\alpha = \bar{\alpha} + h^\top u,$$

where now $\bar{\alpha} \in \mathbb{R}$ and $h \in \mathbb{R}^n$ are parameters to be determined. Such choice of an affine dependence, inspired by the adjustable robust solutions and affine recourse ideas of [5], allows for a reduction of the conservativeness while still resulting in a tractable problem formulation. The inclusion conditions now become, for $i = 1, \ldots, m$,

$$\xi_i^- \leq -2[QA^\top]_i(c + ru) + [Q\omega]_i + \bar{\alpha} + h^\top u \leq \xi_i^+, \ \forall u : \|u\|_2 \leq 1,$$

that is

(4.2) $\quad \xi_i^- \leq -2[QA^\top]_i c + \big(-2r[QA^\top]_i + h^\top\big)u + [Q\omega]_i + \bar{\alpha} \leq \xi_i^+, \ \forall u : \|u\|_2 \leq 1.$

Proceeding as previously by computing explicitly the maximum and the minimum of the central expression in (4.2) with respect to $u$, we obtain that (4.2) holds for $i = 1, \ldots, m$, if and only if

(4.3) $\qquad -2[QA^\top]_i c + \|h^\top - 2r[QA^\top]_i\|_2 + [Q\omega]_i + \bar{\alpha} \leq \xi_i^+,$

(4.4) $\qquad -2[QA^\top]_i c - \|h^\top - 2r[QA^\top]_i\|_2 + [Q\omega]_i + \bar{\alpha} \geq \xi_i^-,$

hold for $i = 1, \ldots, m$. These constitute a set of convex second-order cone (SOC) constraints on the decision variables $c, r, \bar{\alpha}$ and $h$ that guarantee $\mathcal{B}(c, r) \subseteq \mathcal{X}_d$.

Next, we consider the inclusion $\mathcal{B}(c, r) \subseteq \mathcal{H}$. This holds if and only if $\|x - a^{(i)}\|_2 \leq \sqrt{\xi_i^+}$ holds for all $x \in \mathcal{B}(c, r)$ and all $i$, that is if an only if

$$\max_{u : \|u\|_2 \leq 1} \|c - a^{(i)} + ru\|_2 \leq \sqrt{\xi_i^+}, \quad i = 1, \ldots, m.$$

Since by the triangle inequality

$$\|c - a^{(i)} + ru\|_2 \leq \|c - a^{(i)}\|_2 + r\|u\|_2,$$

with equality holding when $u$ is parallel to $c - a^{(i)}$, we have that $\mathcal{B}(c, r) \subseteq \mathcal{H}$ holds if and only if

(4.5) $\qquad \|c - a^{(i)}\|_2 + r \leq \sqrt{\xi_i^+}, \quad i = 1, \ldots, m.$

We have therefore the following

PROPOSITION 4.1. *A ball $\mathcal{B}(c, r) \subseteq \mathcal{X}$ can be computed by solving the following convex SOC program*

(4.6) $\qquad \max_{c \in \mathbb{R}^n, r \geq 0, \bar{\alpha} \in \mathbb{R}, h \in \mathbb{R}^n} r$

$$\text{s.t.} : \ (4.3), (4.4), (4.5), \quad i = 1, \ldots, m.$$

**4.2. Computing an inscribed ellipsoid.** Generalizing the approach described in the previous section, we next discuss how to compute a large (in the sense of the sum of semi-axes lengths) ellipsoid $\mathcal{E}$ inscribed in $\mathcal{X}$. We describe the ellipsoid by

$$\mathcal{E}(c, W) = \{x \in \mathbb{R}^n : x = c + Wu, \|u\|_2 \leq 1\},$$

where $W \in \mathbb{S}^n$ is a positive semidefinite matrix (which we denote by $W \succeq 0$) describing the ellipsoid's shape, and $c \in \mathbb{R}^n$ is the ellipsoid's center. We have in this case (see, e.g., Sec. 10.4.3 in [10]) that $\mathcal{E}(c, W) \subseteq \mathcal{X}_d$ if

$$\xi^- \leq -2QA^\top c - 2QA^\top W u + Q\omega + \alpha(u)\mathbf{1} \leq \xi^+$$

holds for all $u : \|u\|_2 \leq 1$. Posing as before $\alpha = \bar{\alpha} + h^\top u$ and taking the maximum and minimum of the central expression above with respect to $u$ we obtain the equivalent conditions

$$\text{(4.7)} \qquad -2[QA^\top]_i c + \|h^\top - 2[QA^\top]_i W\|_2 + [Q\omega]_i + \bar{\alpha} \leq \xi_i^+$$
$$\text{(4.8)} \qquad -2[QA^\top]_i c - \|h^\top - 2[QA^\top]_i W\|_2 + [Q\omega]_i + \bar{\alpha} \geq \xi_i^-.$$

The inclusion $\mathcal{E}(c, W) \subseteq \mathcal{H}_i$ holds instead if and only if $\|c - a^{(i)} + Wu\|_2 \leq \sqrt{\xi_i^+}$ for all $u : \|u\|_2 \leq 1$. Since

$$\begin{aligned}\|c - a^{(i)} + Wu\|_2 &\leq \|c - a^{(i)}\|_2 + \|Wu\|_2 \\ &\leq \|c - a^{(i)}\|_2 + \|W\|\,\|u\|_2 \\ &\leq \|c - a^{(i)}\|_2 + \|W\|,\end{aligned}$$

where $\|W\|$ denotes the spectral (maximum singular value) norm of $W$, we have that a sufficient (and generally conservative) condition for the inclusion $\mathcal{E}(c, W) \subseteq \mathcal{H}_i$ is given by

$$\|c - a^{(i)}\|_2 + \|W\| \leq \sqrt{\xi_i^+},$$

which in turn can be represented using a slack variable $\gamma \geq 0$ as

$$\text{(4.9)} \qquad \|c - a^{(i)}\|_2 + \gamma \leq \sqrt{\xi_i^+},$$

with $W \preceq \gamma I$. We therefore have the following

PROPOSITION 4.2. *An ellipsoid $\mathcal{E}(c, W) \subseteq \mathcal{X}$ can be computed by solving the following convex SDP problem:*

$$\text{(4.10)} \qquad \begin{aligned}\max_{c \in \mathbb{R}^n, W \in \mathbb{S}^n, \bar{\alpha}, \gamma \in \mathbb{R}, h \in \mathbb{R}^n} &\quad \mathrm{trace}(W) \\ \text{s.t.} : &\quad (4.7), (4.8), (4.9), \quad i = 1, \ldots, m, \\ &\quad 0 \preceq W \preceq \gamma I.\end{aligned}$$

**5. Outer bounding $\mathcal{X}_{\text{true}}$.** The approaches in the previous sections were aimed at finding inner approximations of $\mathcal{X}$. We next describe two techniques aimed at finding outer bounding sets for $\mathcal{X}$, and hence for $\mathcal{X}_{\text{true}}$. The first one in Section 5.1 focuses on finding intervals of containment along a given set of directions, which results in determination of a minimal outer box containing $\mathcal{X}$. The second technique in Section 5.2 discusses how to determine an outer-bounndig ellipsoid for $\mathcal{X}$.

**5.1. Computing an outer-bounding box for $\mathcal{X}$.** Suppose we are given $n$ orthogonal directions $v^{(1)}, \ldots, v^{(n)}$, such that $v^{(i)\top} v^{(j)} = \delta_{ij}$, $\delta_{ij} = 1$ if $i = j$ and $\delta_{ij} = 0$ otherwise. For each $i = 1, \ldots, n$, we compute the maximum and minimum components of $x \in \mathcal{X}$ along the direction $v^{(i)}$, by solving SOCPs

$$
\begin{aligned}
\gamma_i^{\max} = \max_{x \in \mathbb{R}^n, \alpha \in \mathbb{R}} \quad & x^\top v^{(i)} \\
\text{s.t.:} \quad & \xi^- \leq -2QA^\top x + Q\omega + \alpha \mathbf{1} \leq \xi^+, \\
& \|x - a^{(j)}\|_2 \leq \sqrt{\xi_j^+}, \quad j = 1, \ldots, m,
\end{aligned}
\tag{5.1}
$$

$$
\begin{aligned}
\gamma_i^{\min} = \min_{x \in \mathbb{R}^n, \alpha \in \mathbb{R}} \quad & x^\top v^{(i)} \\
\text{s.t.:} \quad & \xi^- \leq -2QA^\top x + Q\omega + \alpha \mathbf{1} \leq \xi^+, \\
& \|x - a^{(j)}\|_2 \leq \sqrt{\xi_j^+}, \quad j = 1, \ldots, m.
\end{aligned}
\tag{5.2}
$$

PROPOSITION 5.1. *Let $\{v^{(1)}, \ldots, v^{(n)}\}$ be an orthonormal set of vectors in $\mathbb{R}^n$, and let $\gamma^{\min}, \gamma^{\max}$ be the vectors with entries $\gamma_i^{\min}, \gamma_i^{\max}$, respectively, computed by solving (5.1), (5.2). Then, the box*

$$
H_{\text{outer}} \doteq \{x \in \mathbb{R}^n : x = \sum_{i=1}^n \gamma_i v^{(i)}, \gamma^{\min} \leq \gamma \leq \gamma^{\max}\}
$$

*is such that $\mathcal{X}_{\text{true}} \subset \mathcal{X} \subseteq H_{\text{outer}}$.*

**Proof.** Take any $x \in \mathcal{X}$. Since $\{v^{(1)}, \ldots, v^{(n)}\}$ form an orthonormal basis for $\mathbb{R}^n$, there exist coefficients $\gamma_i \in \mathbb{R}$ such that $x = \sum_{i=1}^n \gamma_i v^{(i)}$, and these are actually the components of $x$ along the basis vectors, that is $\gamma_i = x^\top v^{(i)}$, $i = 1, \ldots, n$. Now, by definition of $\mathcal{X}$, the selected point $x$ satisfies the constraints of problems (5.1), (5.2), that is, it is feasible albeit possibly not optimal for these problems. Hence, it holds that $\gamma_i^{\min} \leq x^\top v^{(i)} \leq \gamma_i^{\max}$, for $i = 1, \ldots, n$, that is $\gamma_i^{\min} \leq \gamma_i \leq \gamma_i^{\max}$, for $i = 1, \ldots, n$, which means that $x \in H_{\text{outer}}$, which proves the inclusion $\mathcal{X} \subseteq H_{\text{outer}}$. □

For example, the $v^{(i)}$s can be taken to be the standard basis of $\mathbb{R}^n$, in which case we obtain an hyperrectangle (i.e., a box with sides aligned to the coordinate axes). Another choice would be to take as $v^{(i)}$s the set of orthogonal eigenvectors of the $W$ matrix of the inscribed ellipsoid $\mathcal{E}$ of Proposition 4.2, which describe the directions of the semiaxes; we find in this way the minimal box with sides aligned to the ellipsoid axes that contains $\mathcal{X}$, $\mathcal{E}$ and $\mathcal{X}_{\text{true}}$.

REMARK 5 (Complexity of box outer bounding). The typical situations of practical interest in localization involve small $n$ (say, $n = 2$, or $n = 3$) and a moderate number of anchors $m > n$. Typical instances of SOCP problems (5.1) and (5.2), therefore, do not require solution methods tailored for large-scale optimization, and can be solved effectively by means of standard primal-dual interior-point algorithms. Each of these problems involve $n + 1$ variables, $m$ SOC constraints and $2m$ linear inequality constraints. According to the classical self-concordant barrier complexity framework of [20] for interior-point algorithms, the barrier parameter for these problems is given by $\nu = 4m$, hence the outer-iteration count to reach $\epsilon$-accuracy is $O(\sqrt{\nu} \log(1/\epsilon)) = O(2\sqrt{m} \log(1/\epsilon))$. Each iteration in turn requires roughly $O(mn^3 + 2m^2n^2 + 8m^3)$ floating-point operations. The computational effort for computing the bounding box thus remains moderate, under the typical conditions of interest in practical localization applications. ⋆

**5.1.1. Directional localization.** As a variant of the previous problem, we can also determine a guaranteed uncertainty range for the localization along any desired direction. Suppose we fix a probing direction $v \in \mathbb{R}^n$, $\|v\|_2 = 1$, and we ask ourselves what is the range of uncertainty in $x$ around a central estimate $c \in \mathcal{X}$, along direction $v$. In formulae, we consider $x = c + \gamma v$, with $c$ and $v$ given, and we seek for the maximum and minimum values of $\gamma$ for which $x \in \mathcal{X}$. These values can be computed exactly by solving two SOCPs, as follows

$$\begin{aligned}
\gamma_{\max} = \max_{\gamma, \alpha \in \mathbb{R}} \quad & \gamma \\
\text{s.t.} : \quad & \xi^- \leq -2QA^\top(c + \gamma v) + Q\omega + \alpha \mathbf{1} \leq \xi^+, \\
& \|c + \gamma v - a^{(i)}\|_2 \leq \sqrt{\xi_i^+}, \quad i = 1, \ldots, m,
\end{aligned} \tag{5.3}$$

$$\begin{aligned}
\gamma_{\min} = \min_{\gamma, \alpha \in \mathbb{R}} \quad & \gamma \\
\text{s.t.} : \quad & \xi^- \leq -2QA^\top(c + \gamma v) + Q\omega + \alpha \mathbf{1} \leq \xi^+, \\
& \|c + \gamma v - a^{(i)}\|_2 \leq \sqrt{\xi_i^+}, \quad i = 1, \ldots, m.
\end{aligned} \tag{5.4}$$

**5.2. Computing an outer-bounding ellipsoid for $\mathcal{X}$.** Consider an ellipsoid $\mathcal{O}(c, P)$ described by the equation

$$\mathcal{O}(c, P) = \{x \in \mathbb{R}^n : (x - c)^\top P^{-1}(x - c) \leq 1\},$$

where $P \in \mathbb{S}^n$ is a positive definite matrix describing the ellipsoid's shape, and $c \in \mathbb{R}^n$ is the ellipsoid's center. Consider further the representation of the polyhedron $\mathcal{X}_d$ given in Proposition 3.1

$$\mathcal{X}_d \doteq \{x \in \mathbb{R}^n : 2EA^\top x = E(\omega - \xi), \xi \in H\},$$

where $H$ is the hyperrectangle in $\mathbb{R}^m$ defined in (2.8), and we assume the standard conditions of Remark 4 to hold, so that $EA^\top$ is full-column rank and hence $\mathcal{X}_d$ is bounded, i.e., it is a polytope. $H$ has $2^m$ vertices $v^{(1)}, \ldots, v^{(q)}$, $q = 2^m$, and $\mathcal{X}_d$ can be represented as the convex hull of the images $w^{(i)}$ of $v^{(i)}$, $i = 1, \ldots, q$, through the linear map

$$w^{(i)} = \frac{1}{2}(EA^\top)^\dagger E(\omega - v^{(i)}), \quad i = 1, \ldots, q. \tag{5.5}$$

The inclusion $\mathcal{O} \supseteq \mathcal{X}_d$ is therefore satisfied if $w^{(i)} \in \mathcal{O}$, $i = 1, \ldots, q$, that is if

$$(w^{(i)} - c)^\top P^{-1}(w^{(i)} - c) \leq 1, \quad i = 1, \ldots, q. \tag{5.6}$$

For the inclusion $\mathcal{O} \supseteq \mathcal{H}$, where $\mathcal{H} \doteq \bigcap_{i=1}^m \mathcal{H}_i$ with $\mathcal{H}_i$ being the balls defined in (3.6), we use a standard sufficient condition for computing an ellipsoid covering the intersection of ellipsoids (closed balls are a special case of ellipsoids) given by the $\mathcal{S}$-procedure, see, e.g., Chapter 3 of [8] or Section 3.2 in [25]. The following proposition holds.

PROPOSITION 5.2. *A suboptimal minimum-size ellipsoid $\mathcal{O}(c, P) \supseteq \mathcal{X}$ can be found by solving for $P \in \mathbb{S}^n$, $c \in \mathbb{R}^n$ and $\tau \in \mathbb{R}^m$ the semidefinite optimization problem (SDP)*

$$
\begin{aligned}
\min_{P, c, \tau \geq 0} \quad & \operatorname{trace}(P) \\
\text{s.t.:} \quad & \begin{bmatrix} 1 & (w^{(i)} - c)^\top \\ * & P \end{bmatrix} \succeq 0, \quad i = 1, \ldots, 2^m, \\
& \begin{bmatrix} \sum_{i=1}^m \tau_i I_n & -\sum_{i=1}^m \tau_i a^{(i)} & -I_n \\ * & 1 + \sum_{i=1}^m \tau_i(\|a^{(i)}\|_2^2 - \xi_i^+) & -c^\top \\ * & * & P \end{bmatrix} \succeq 0.
\end{aligned}
$$
(5.7)

**Proof.** The first set of conditions is obtained from (5.6) via Schur complement, and implies $\mathcal{O} \supseteq \mathcal{X}_d$. The second linear matrix inequality condition derives from application of the $\mathcal{S}$-procedure to the implication

$$x^\top x - 2a^{(i)\top} x + \|a^{(i)}\|_2^2 - \xi_i^+ \leq 0, \ i = 1, \ldots, m \ \Rightarrow \ (x - c)^\top P^{-1}(x - c) - 1 \leq 0,$$

via the approach detailed in Section 11.4.2.5 of [10]. This second linear matrix inequality implies $\mathcal{O} \supseteq \mathcal{H}$, and therefore conditions in problem (5.7) jointly enforce $\mathcal{O} \supseteq \mathcal{X}_d \cap \mathcal{H} = \mathcal{X}$. □

REMARK 6 (Complexity of ellipsoidal outer bounding). Problem (5.7) is a standard convex SDP which can be solved to global optimality via primal-dual interior-point methods. The problem involves $q = 2^m$ LMI blocks of size $(n+1) \times (n+1)$ and one LMI block of size $(2n+1) \times (2n+1)$. According to the classical self-concordant barrier framework of [20], the outer-iteration count to reach $\epsilon$-accuracy is $O(\sqrt{\nu} \log(1/\epsilon))$, where $\nu = q(n+1) + (2n+1)$, whence the number of outer iterations scales as

$$O(\sqrt{q(n+1) + (2n+1)} \log(1/\epsilon)),$$

while each iteration essentially requires $O(q(n+1)^3 + (2n+1)^3)$ floating-point operations. Therefore, the computational cost grows linearly in $q$ and cubically in the LMI blocks' sizes. However, due to the exponential dependence $q = 2^m$, the computation of an ellipsoidal outer bound is more demanding than the computation of a box outer bound (cfr. the complexity analysis in Remark 5), and may become impractical when the number $m$ of measurements increases. ⋆

REMARK 7 (Central location estimate). We can use the center $c_e$ of the inscribed ellipsoid computed in Proposition 4.2, or the center $c_s$ of the inscribed ball computed in Proposition 4.1, as a central location estimate. Also, we can obtain central estimates via the outer bounding sets, e.g., via the center of the outer-bounding box $H_{\text{outer}}$ given in Proposition 5.1, or via the center of the outer-bounding ellipsoid $\mathcal{O}$ given in Proposition 5.2. The centers $c_e$ and $c_s$ of the inscribed sets belong to $\mathcal{X}$ by construction, although they are in general *not* guaranteed to belong to $\mathcal{X}_{\text{true}}$. ⋆

**5.3. Algorithmic summary of the localization procedure.** For clarity and ease of implementation, we summarize here the main computational steps required to obtain the set-membership outer-bounding localization estimates from the raw measurements.
  1. **Input data:** Collect anchor positions $A = [a^{(1)} \cdots a^{(m)}]$ and the $m$ range measurements (e.g., $y_i$ or $z_i$). Set the measurement bounds $\xi^-, \xi^+$ according to the assumed absolute or relative error model (see Section 2).

2. **Pre-processing & Outlier check (optional):** To safeguard against bound violations, solve the convex program (6.6) to find the minimal bound enlargements $\nu^+, \nu^-$. Update $\xi^- \leftarrow \xi^- - \nu^-$ and $\xi^+ \leftarrow \xi^+ + \nu^+$ (see Section 6.2 below).
3. **Construct constants:** Compute matrix $E$ (the incidence matrix of the difference-of-measurements graph, Section 3.1), vector $\omega$, and projector $Q = \frac{1}{m}(mI_m - \mathbf{1}_m\mathbf{1}_m^\top)$.
4. **Outer Bounding (Box):** Choose $n$ orthogonal directions $v^{(i)}$ (e.g., the standard basis). For each direction, solve the SOCPs (5.1) and (5.2) to find the bounds $\gamma_i^{\max}$ and $\gamma_i^{\min}$. The Cartesian product of these intervals yields the guaranteed outer bounding box $H_{\text{outer}}$ (Proposition 5.1).
5. **Inner Approximation (Ellipsoid):** Solve the SDP (4.10) to find the maximal inscribed ellipsoid $\mathcal{E}(c, W)$ (Proposition 4.2), using the pre-computed bounds and matrices.

**6. Additional insights and results.** In this section we provide some additional insights into the set-membership localization problem, examining in particular a technique for trying to find a point belonging to $\mathcal{X}_{\text{true}}$, and a discussion on how to deal with situations in which the measurement errors violate the assumed upper or lower bounds.

**6.1. Searching locally for a feasible point.** As observed in Remark 7, the central estimate $c_e$ given by the center of the inscribed ellipsoid $\mathcal{E} \subset \mathcal{X}$ may fail to belong to $\mathcal{X}_{\text{true}}$, and the same can happen for any other of the central estimates discussed in the previous sections, such as the center $c_b$ of the outer bounding box $H_{\text{outer}}$. In practice, this should not be an issue, since we are mostly interested in locating $\mathcal{X}_{\text{true}}$ via bounding sets, hence the center of such sets still carries the key information about where approximately $\mathcal{X}_{\text{true}}$ is located. However, if one insists in trying to find a point which truly belongs to $\mathcal{X}_{\text{true}}$, the following *local* approach can be followed.

We let $c$ denote a central estimate, computed as discussed in the previous sections; for instance, we may take $c$ to be the center of the outer bounding box $H_{\text{outer}} \supseteq \mathcal{X}_{\text{true}}$. If $c \in \mathcal{X}_{\text{true}}$ (this condition is easy to check numerically), then there is nothing else to do. If instead $c \notin \mathcal{X}_{\text{true}}$, we search locally for a possible point that belongs to $\mathcal{X}_{\text{true}}$. The key observation is that when dealing with the original nonlinear constraints (2.9) the right hand side inequality poses no problem from a computational point of view, since it represents a convex constraint on $x$. Contrary, the left-hand side inequality is nonconvex in $x$. The idea we propose is that once a neighborhood of $\mathcal{X}_{\text{true}}$, and a corresponding center $c$, has been determined, we can approximate locally the nonconvex constraint $\|x - a^{(i)}\|_2^2 \geq \xi_i^-$ by means of its first-order approximation around point

$$c^{(i)} \doteq \sqrt{\xi_i^-} \frac{c - a^{(i)}}{\|c - a^{(i)}\|_2},$$

that is, by means of an affine inequality of the form

(6.1) $$2(c^{(i)} - a^{(i)})^\top (x - c^{(i)}) \geq 0.$$

Figure 4 illustrates this idea on a two dimensional example. Intuitively, such approximation works best in a "far field" regime in which $c$ is far from the anchors. One approach is thus to compute a new point $x$ that belongs to the intersection of $\mathcal{X}$ and the regions where (6.1) hold for $i = 1, \ldots, m$. In practice, since the constraints (6.1) are locally more stringent than the original ones, to avoid infeasibility we add slack variables $s_i$ to allow some constraint violation, and we penalize such violations in the objective. Overall, we solve a convex optimization problem of the form

$$
\begin{aligned}
\min_{x \in \mathbb{R}^n, s \in \mathbb{R}^m, \alpha \in \mathbb{R}} \quad & \sum_{i=1}^m |s_i| \\
\text{s.t.:} \quad & \xi^- \leq Q\omega - 2QA^\top x + \alpha \mathbf{1} \leq \xi^+ \\
& \|x - a^{(i)}\|_2^2 \leq \xi_i^+, \quad i = 1, \ldots, m, \\
& 2(c^{(i)} - a^{(i)})^\top (x - c^{(i)}) + s_i \geq 0, \quad i = 1, \ldots, m.
\end{aligned}
\tag{6.2}
$$

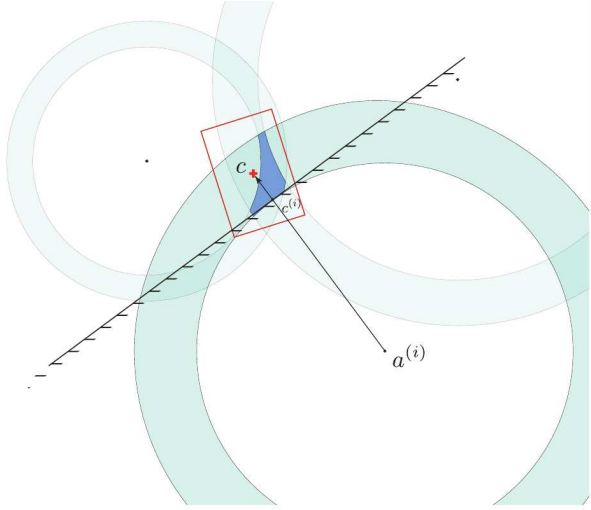

Fig. 4. *Local linear approximation of the nonconvex constraint $\|x - a^{(i)}\|_2^2 \geq \xi_i^-$. The red box represents $H_{\text{outer}}$, with center $c$ which, in this illustration, does not belong to $\mathcal{X}_{\text{true}}$ (the darker shaded area).*

The above problem is a convex quadratic programming problem (QP) which can be solved globally and with great efficiency. We remark, however, that the linearized constraints are still only an approximation of the actual nonlinear constraints, whence the optimal solution $\hat{x}$ that we obtain from (6.2) may still fail to belong to $\mathcal{X}_{\text{true}}$.

**6.2. Errors that violate their bounds.** As discussed in Remark 3, when the errors $\epsilon_i$ in the measurement equations (2.1) – resp., the errors $e_i$ in the measurement equations (2.2) – respect their assumed bounds, then the corresponding $\xi_i$ parameters are within the bounds $\xi_i^-, \xi_i^+$, and the set $\mathcal{X}_{\text{true}}$ is nonempty. Contrary, when some of the actual errors happen to be outside their assumed bounds, the sets $\mathcal{S}_i$ in (2.9) may have empty intersection, hence $\mathcal{X}_{\text{true}}$, $\mathcal{X}_d$ and $\mathcal{X}$ may be empty. In such case, all the convex optimization problems discussed in the previous sections may turn out to be infeasible.

The described approach, however, can be modified so to accommodate the practical situation in which some measurements may be affected by outliers or, in general, by errors that do not respect the assumed bounds. The idea is simply to allow for an enlargement of the bounds $\xi^-, \xi^+$. That is, we consider *variable* bounds $\xi^- - \nu^-, \xi^+ + \nu^+$, where $\nu^+, \nu^- \geq 0$ contain the upper and lower bound enlargements, respectively, and we seek for the minimum enlargement which guarantees that a ball of given radius $r \geq 0$ is contained in the new, enlarged, localization set $\mathcal{X}'$. By looking at the derivations in Section 4.1 we see that a ball $\mathcal{B}(c, r)$ is contained in $\mathcal{X}'$ if

$$
\begin{align}
-2[QA^\top]_i c + \|h^\top - 2r[QA^\top]_i\|_2 + [Q\omega]_i + \bar{\alpha} &\leq \xi_i^+ + \nu^+, \tag{6.3}\\
-2[QA^\top]_i c - \|h^\top - 2r[QA^\top]_i\|_2 + [Q\omega]_i + \bar{\alpha} &\geq \xi_i^- - \nu^-, \tag{6.4}
\end{align}
$$

hold for $i = 1, \ldots, m$, and

$$
\|c - a^{(i)}\|_2^2 + 2r\|c - a^{(i)}\|_2 + r^2 \leq \xi_i^+ + \nu_i^+, \tag{6.5}
$$

hold for $i = 1, \ldots, m$, where this latter inequality is obtained by squaring (4.5). For given and typically small $r \geq 0$ we solve the convex program

$$
\begin{align}
(6.6) \quad \min_{c \in \mathbb{R}^n, h \in \mathbb{R}^n, \nu^+, \nu^- \in \mathbb{R}^m, \bar{\alpha} \in \mathbb{R}} \quad & \sum_{j=1}^m (\nu_j^+ + \nu_j^-)\\
\text{s.t.} : \quad & \nu^+, \nu^- \geq 0, (6.3), (6.4), (6.5), \quad i = 1, \ldots, m.
\end{align}
$$

This problem is always feasible. If at the optimum $\nu^+, \nu^- = 0$, then the original localization set $\mathcal{X}$ was nonempty. If instead we have $\nu^+ \geq 0$, $\nu^+ \neq 0$, ans/or $\nu^- \geq 0$, $\nu^- \neq 0$ at optimum, then $\mathcal{X}$ was empty, but we can enlarge the error bounds as $\xi^- \leftarrow \xi^- - \nu^-$, $\xi^+ \leftarrow \xi^+ + \nu^+$ and be guaranteed that the corresponding updated set $\mathcal{X}'$ is nonempty and contains a ball of radius $r$. Vectors $\nu^-$ or $\nu^+$ with positive entries thus signal that the localization set is empty due to some of the measurements not respecting their assumed bounds. When this happens, we can inspect the size of the positive elements in $\nu^+, \nu^-$ relative to the corresponding baseline values in $\xi^+, \xi^-$, and decide whether we want to discard this experiment and repeat it with new measurements, or if we want to proceed to location estimation via the standard approach applied using the updated enlarged bounds. Problem (6.6) can thus be solved as a pre-processing step, before running the actual localization, for validating the feasibility of the assumed error bounds, for removing outliers, or for adjusting the error bounds so that the measurement equations become jointly feasible.

**7. Numerical experiments.** In these numerical experiments, we consider the measurement model (2.2) with relative additive errors $\tilde{e}_i \in [\tilde{e}_i^-, \tilde{e}_i^+]$, and we take $\tilde{e}_i^- = -\tilde{e}_i^+ < 0$ for all $i$; the corresponding $\xi_i$s and the relative bounds are given by (2.7). We experiment with space dimension $n = 2$ and $n = 3$. For given $n$, we select integer $m \in \{n+1, \ldots, 10\}$ and assign the $m$ anchor positions $a^{(i)}$, $i = 1, \ldots, m$, uniformly at random in $[-1000, 1000]^n$. The true object position $x_{\text{true}} \in \mathbb{R}^n$ is placed uniformly at random in $[-100, 100]^n$, and then $m$ noisy measurements $z_i$ are generated as $z_i = (1 + \tilde{e}_i)\|x_{\text{true}} - a^{(i)}\|_2$, $i = 1, \ldots, m$, where $\tilde{e}_i$ are generated uniformly at random in $[-\tilde{e}_i^+, \tilde{e}_i^+]$.

For each random problem instance, we compute the inscribed ellipsoid $\mathcal{E}(c_e, P)$ according to Proposition 4.2 and the bounding box $H_{\text{outer}}$ according to Proposition 5.1, with sides aligned to the ellipsoid $\mathcal{E}$'s axes. We then check if the center $c_b$ of $H_{\text{outer}}$ is in $\mathcal{X}_{\text{true}}$; in the positive case we return as central estimate $\hat{x} = c_b$ and in the negative case we solve problem (6.2) and return the corresponding optimal solution as estimate $\hat{x}$. We further compute the actual distance from $\hat{x}$ to the true object location $x_{\text{true}}$, which represents the absolute position estimation error, and also evaluate the relative error by dividing $\|x_{\text{true}} - \hat{x}\|_2$ by the average distance of $\hat{x}$ from the anchors. Finally, we assess the set-membership localization error as the average of the half-side lengths of $H_{\text{outer}}$.

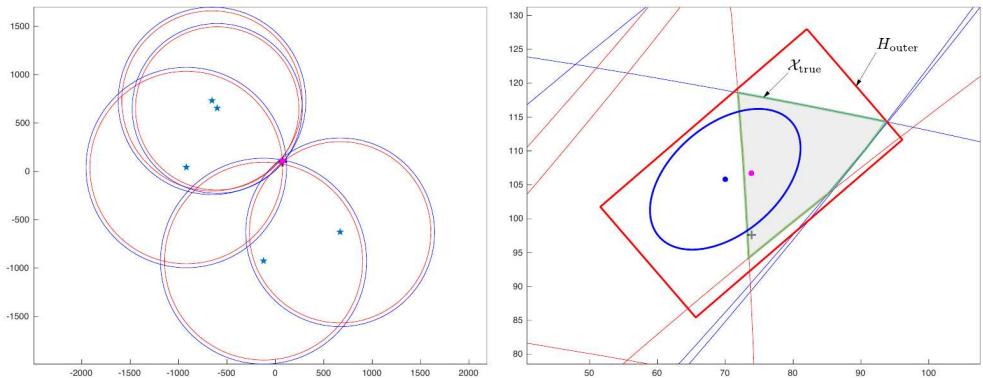

FIG. 5. *Left: five anchors and intersection of their annuli. Right: outer bounding box for $\mathcal{X}_{\text{true}}$ (in red) and inner ellipsoidal approximation of $\mathcal{X} \supset \mathcal{X}_{\text{true}}$ (in blue).*

Figure 5 shows the result of one random instance of the setup described above, with $n = 2$, $m = 5$, and error bound $\tilde{e}_i^+ = 0.02$ (i.e., 2% relative error) for all $i$. In the left panel we see the anchors (stars) and their measurement annuli. The right panel shows a close up of the intersection region: the outer bounding box $H_{\text{outer}}$ is depicted in red, the inner ellipsoid $\mathcal{E}$ is depicted in blue. The black cross indicates $x_{\text{true}}$, and the red dot indicates $c_b$, which in this case does belong to $\mathcal{X}_{\text{true}}$, hence $\hat{x} = c_b$.

In the numerical experiments, for each $n \in \{2,3\}$ we selected in succession $m \in \{n+1, \ldots, 10\}$ and, for each $(n,m)$ pair, we generated and solved $N = 100$ random experiments. We repeated all the experiments three times, with increasing relative error bounds $\tilde{e}_i^+ = 0.02$, $\tilde{e}_i^+ = 0.1$, and $\tilde{e}_i^+ = 0.2$. The test was coded in Matlab, using CVX [12] with Mosek solver [1] for SDP and SOCP, and run on a iMac 3.2 GHz Intel Xeon desktop computer, without any particular care devoted to code optimization.

In each run we evaluated: the absolute error with respect to the box center, $e(c_b) \doteq \|c_b - x_{\text{true}}\|_2$; the absolute error with respect to the modified center $\hat{x}$, $e(\hat{x}) = \|\hat{x} - x_{\text{true}}\|_2$; the relative error $e_\%(c_b)$ obtained by dividing $e(c_b)$ by the average distance of $c_b$ from the anchors; the relative error $e_\%(\hat{x})$ obtained by dividing $e(\hat{x})$ by the average distance of $\hat{x}$ from the anchors; the average $e_{\text{outer}}$ of the half-side lengths of the bounding box $H_{\text{outer}}$; the relative value $e_{\text{outer}\%}$ obtained by dividing $e_{\text{outer}}$ by the average distance of $c_b$ from the anchors; a 0/1 indicator $I_{\mathcal{X}_{\text{true}}}(c_b)$ which is 1 if $c_b \in \mathcal{X}_{\text{true}}$ and zero otherwise; and an analogous indicator $I_{\mathcal{X}_{\text{true}}}(\hat{x})$ for $\hat{x}$. These values are then averaged over the $N$ random experiments; the average of $I_{\mathcal{X}_{\text{true}}}(c_b)$ corresponds to the frequency $f_\%(c_b)$ with which $c_b \in \mathcal{X}_{\text{true}}$, the average of $I_{\mathcal{X}_{\text{true}}}(\hat{x})$ corresponds to the frequency $f_\%(\hat{x})$ with which $\hat{x} \in \mathcal{X}_{\text{true}}$. Further, we evaluated the runtime $t_r$ (in second) needed for computing the bounding box, and we assessed the conservativeness of the inclusion $\mathcal{X} \supseteq \mathcal{X}_{\text{true}}$ by estimating via a Monte-Carlo method the ratio of volumes $\eta_{\text{vol}} = \text{vol}(\mathcal{X})/\text{vol}(\mathcal{X}_{\text{true}})$, obtained by generating $N_x = 10^4$ uniform random points in $\mathcal{X}$ and counting how many of them actually fall in $\mathcal{X}_{\text{true}}$; calling $N_t$ this number, the Monte-Carlo estimate of $\eta_{\text{vol}}$ is simply $N_x/N_t$.

The results for $n = 2$ and $n = 3$ are reported in Table 1 and Table 2, respectively, for relative error level $\tilde{e}_i^+ = 0.02$. The experiments were then repeated for higher error level $\tilde{e}_i^+ = 0.1$, and the corresponding results are reported in Table 3 and Table 4, and further for $\tilde{e}_i^+ = 0.2$, with corresponding results reported in Table 5 and Table 6.

| $m$ | $e(c_b)$ | $e(\hat{x})$ | $e_\%(c_b)$ | $e_\%(\hat{x})$ | $e_{\text{outer}}$ | $e_{\text{outer}\%}$ | $f_\%(c_b)$ | $f_\%(\hat{x})$ | $t_r$ | $\eta_{\text{vol}}$ |
|---|---|---|---|---|---|---|---|---|---|---|
| 3 | 47.53 | 21.38 | 20.16 | 2.45 | 43.25 | 14.88 | 49 | 80 | 0.66 | 2.09 |
| 4 | 17.04 | 14.59 | 1.99 | 1.71 | 22.13 | 2.59 | 58 | 81 | 0.72 | 2.10 |
| 5 | 10.20 | 9.60 | 1.25 | 1.16 | 15.21 | 1.86 | 71 | 85 | 0.79 | 2.11 |
| 6 | 9.08 | 9.94 | 1.09 | 1.21 | 13.47 | 1.64 | 63 | 80 | 0.88 | 2.51 |
| 7 | 7.15 | 7.80 | 0.87 | 0.94 | 9.99 | 1.22 | 67 | 80 | 0.92 | 3.38 |
| 8 | 5.51 | 6.11 | 0.71 | 0.79 | 8.84 | 1.12 | 71 | 76 | 0.97 | 2.83 |
| 9 | 5.36 | 7.17 | 0.65 | 0.85 | 7.87 | 0.96 | 79 | 87 | 1.06 | 2.62 |
| 10 | 4.25 | 4.59 | 0.52 | 0.57 | 6.97 | 0.87 | 74 | 85 | 1.14 | 2.96 |

TABLE 1
*Average results over $N = 100$ randomized tests, for $n = 2$ and relative error level $\tilde{e}_i^+ = 0.02$.*

| $m$ | $e(c_b)$ | $e(\hat{x})$ | $e_\%(c_b)$ | $e_\%(\hat{x})$ | $e_{\text{outer}}$ | $e_{\text{outer}\%}$ | $f_\%(c_b)$ | $f_\%(\hat{x})$ | $t_r$ | $\eta_{\text{vol}}$ |
|---|---|---|---|---|---|---|---|---|---|---|
| 4 | 70.15 | 38.04 | 7.17 | 3.45 | 63.22 | 6.29 | 46 | 87 | 1.13 | 2.55 |
| 5 | 35.08 | 27.54 | 3.30 | 2.59 | 39.43 | 3.75 | 58 | 84 | 1.22 | 2.36 |
| 6 | 24.21 | 25.86 | 2.40 | 2.54 | 29.81 | 2.95 | 49 | 73 | 1.36 | 3.21 |
| 7 | 17.39 | 19.70 | 1.72 | 1.95 | 23.63 | 2.34 | 68 | 86 | 1.46 | 2.59 |
| 8 | 13.57 | 15.95 | 1.31 | 1.55 | 20.91 | 2.02 | 80 | 86 | 1.51 | 2.23 |
| 9 | 11.75 | 14.97 | 1.20 | 1.51 | 17.71 | 1.79 | 73 | 82 | 1.59 | 2.86 |
| 10 | 11.46 | 11.47 | 1.12 | 1.13 | 15.72 | 1.55 | 79 | 87 | 1.68 | 2.89 |

TABLE 2
*Average results over $N = 100$ randomized tests, for $n = 3$ and relative error level $\tilde{e}_i^+ = 0.02$.*

| $m$ | $e(c_b)$ | $e(\hat{x})$ | $e_\%(c_b)$ | $e_\%(\hat{x})$ | $e_{\text{outer}}$ | $e_{\text{outer}\%}$ | $f_\%(c_b)$ | $f_\%(\hat{x})$ | $t_r$ | $\eta_{\text{vol}}$ |
|---|---|---|---|---|---|---|---|---|---|---|
| 3 | 152.49 | 113.21 | 20.55 | 12.52 | 169.84 | 22.47 | 50 | 84 | 0.66 | 2.01 |
| 4 | 94.12 | 82.45 | 12.70 | 10.02 | 115.30 | 15.12 | 48 | 79 | 0.72 | 2.46 |
| 5 | 56.31 | 61.17 | 7.22 | 7.41 | 79.88 | 10.14 | 62 | 86 | 0.80 | 2.16 |
| 6 | 38.83 | 43.18 | 4.95 | 5.38 | 60.87 | 7.73 | 71 | 89 | 0.84 | 2.22 |
| 7 | 37.49 | 40.71 | 4.92 | 5.14 | 56.76 | 7.29 | 63 | 75 | 0.92 | 2.44 |
| 8 | 29.35 | 36.86 | 3.70 | 4.55 | 45.35 | 5.70 | 62 | 75 | 1.00 | 2.88 |
| 9 | 24.98 | 29.67 | 3.16 | 3.71 | 41.38 | 5.17 | 73 | 83 | 1.16 | 2.93 |
| 10 | 22.97 | 25.81 | 2.85 | 3.20 | 39.07 | 4.84 | 74 | 87 | 1.16 | 2.75 |

TABLE 3
*Average results over $N = 100$ randomized tests, for $n = 2$ and relative error level $\tilde{e}_i^+ = 0.1$.*

We observe, as expected, that the absolute and relative errors decrease as the number $m$ of anchors (and hence of measurements) increases. Also, we see that the local search step that computes $\hat{x}$ consistently improves the probability that the resulting estimate actually belongs to $\mathcal{X}_{\text{true}}$, compared to the plain outer box center $c_b$. However, the absolute and relative errors with $\hat{x}$ improve compared to $c_b$ only for low values of $m$ approximately, up to $m = 4$ for $n = 2$, and up to $m = 6$ for $n = 3$), while as $m$ increases there seems to be no advantage in using $\hat{x}$ in place of $c_b$ in terms of absolute or relative localization errors. The average time $t_r$ for computing an outer box remains below 1.16s for $n = 2$ and below 1.72s for $n = 3$; it mildly grows with $n$ and $m$, but it is insensitive to the noise level. We further observe that

| $m$ | $e(c_b)$ | $e(\hat{x})$ | $e_\%(c_b)$ | $e_\%(\hat{x})$ | $e_\text{outer}$ | $e_{\text{outer}\%}$ | $f_\%(c_b)$ | $f_\%(\hat{x})$ | $t_r$ | $\eta_\text{vol}$ |
|---|---|---|---|---|---|---|---|---|---|---|
| 4 | 235.29 | 181.87 | 24.89 | 16.22 | 248.16 | 26.01 | 44 | 92 | 1.12 | 1.96 |
| 5 | 154.53 | 138.54 | 15.65 | 12.88 | 179.08 | 18.09 | 56 | 88 | 1.20 | 2.08 |
| 6 | 113.20 | 117.72 | 11.47 | 11.40 | 140.00 | 14.10 | 63 | 89 | 1.28 | 2.27 |
| 7 | 85.60 | 94.70 | 8.47 | 9.02 | 119.11 | 11.77 | 60 | 76 | 1.35 | 2.28 |
| 8 | 78.80 | 93.76 | 7.80 | 9.04 | 108.14 | 10.65 | 66 | 82 | 1.52 | 2.59 |
| 9 | 66.49 | 71.77 | 6.80 | 7.15 | 89.20 | 9.11 | 74 | 86 | 1.54 | 2.79 |
| 10 | 59.44 | 73.75 | 5.88 | 7.19 | 82.48 | 8.13 | 70 | 82 | 1.66 | 2.92 |

TABLE 4

Average results over $N=100$ randomized tests, for $n=3$ and relative error level $\tilde{e}_i^+ = 0.1$.

| $m$ | $e(c_b)$ | $e(\hat{x})$ | $e_\%(c_b)$ | $e_\%(\hat{x})$ | $e_\text{outer}$ | $e_{\text{outer}\%}$ | $f_\%(c_b)$ | $f_\%(\hat{x})$ | $t_r$ | $\eta_\text{vol}$ |
|---|---|---|---|---|---|---|---|---|---|---|
| 3 | 238.37 | 214.46 | 36.01 | 22.71 | 321.63 | 45.87 | 49 | 91 | 0.67 | 1.71 |
| 4 | 142.12 | 150.90 | 17.92 | 16.71 | 198.32 | 24.55 | 54 | 90 | 0.75 | 1.63 |
| 5 | 111.17 | 119.91 | 14.24 | 13.62 | 168.60 | 21.18 | 54 | 82 | 0.86 | 2.21 |
| 6 | 90.30 | 105.34 | 11.37 | 12.16 | 136.06 | 17.00 | 56 | 82 | 0.92 | 2.31 |
| 7 | 74.45 | 83.57 | 9.30 | 10.14 | 107.81 | 13.50 | 75 | 88 | 0.97 | 2.27 |
| 8 | 62.41 | 73.44 | 7.78 | 9.03 | 95.68 | 11.89 | 74 | 81 | 1.01 | 2.31 |
| 9 | 58.01 | 67.40 | 7.22 | 8.10 | 88.57 | 10.93 | 69 | 83 | 1.07 | 2.75 |
| 10 | 48.81 | 54.17 | 6.19 | 6.74 | 78.18 | 9.84 | 66 | 76 | 1.13 | 3.39 |

TABLE 5

Average results over $N=100$ randomized tests, for $n=2$ and relative error level $\tilde{e}_i^+ = 0.2$.

| $m$ | $e(c_b)$ | $e(\hat{x})$ | $e_\%(c_b)$ | $e_\%(\hat{x})$ | $e_\text{outer}$ | $e_{\text{outer}\%}$ | $f_\%(c_b)$ | $f_\%(\hat{x})$ | $t_r$ | $\eta_\text{vol}$ |
|---|---|---|---|---|---|---|---|---|---|---|
| 4 | 403.80 | 347.49 | 46.31 | 28.74 | 469.07 | 52.50 | 32 | 91 | 1.21 | 1.77 |
| 5 | 234.85 | 237.01 | 24.61 | 21.15 | 328.62 | 33.67 | 63 | 93 | 1.31 | 1.64 |
| 6 | 219.68 | 226.07 | 22.32 | 20.60 | 302.67 | 30.32 | 57 | 92 | 1.41 | 1.99 |
| 7 | 193.36 | 192.95 | 19.75 | 17.88 | 257.53 | 26.03 | 48 | 86 | 1.42 | 2.96 |
| 8 | 145.29 | 170.62 | 14.52 | 15.90 | 213.12 | 21.21 | 66 | 86 | 1.51 | 2.54 |
| 9 | 126.69 | 158.13 | 12.60 | 15.25 | 178.07 | 17.75 | 71 | 86 | 1.65 | 2.29 |
| 10 | 122.60 | 145.87 | 12.12 | 13.73 | 175.43 | 17.39 | 68 | 82 | 1.72 | 3.25 |

TABLE 6

Average results over $N=100$ randomized tests, for $n=3$ and relative error level $\tilde{e}_i^+ = 0.2$.

the conservativeness of the inclusion $\mathcal{X}_\text{true} \subseteq \mathcal{X}$, as measured by the Monte-Carlo estimated volume ratio $\eta_\text{vol}$, seems to moderately increase with the number $m$ of measurements while being little influenced by $n$.

Finally, we tested the proposed set-membership estimation technique in the situation where the measurements may be affected by outliers, that is by measurement errors that violate the assumed bounds. We considered the case of $n=2$ and $m \in \{3, \ldots, 10\}$. We assumed 2% noise bounds, i.e., $\tilde{e}_i^+ = 0.02$, but the actual measurement noise was generated in the simulation as

$$z_i = \begin{cases} \tilde{e}_i^+(1+u_i)\|x_\text{true} - a^{(i)}\|_2 & \text{with probability } 0.9 \\ 0.5(1+u_i)\|x_\text{true} - a^{(i)}\|_2 & \text{with probability } 0.1, \end{cases}$$

where $u_i$ are generated uniformly at random in $[-1, 1]$. This means that the measurement errors respect the 2% assumed bound with probability 0.9, while with probability 0.1 they unexpectedly violate the assumed bound and may actually have a much larger level of relative error (up to 50% error). For each pair $(n, m)$ we generated $N = 100$ random problem instances and, for each instance, we first solved Problem (6.6), assuming a feasibility radius $r = 0.01$. The solution to this problem yielded, whenever necessary, the minimal noise bound enlargements $\nu^+, \nu^-$ that guarantee that the updated bounds $\xi^- \leftarrow \xi^- - \nu^-$, $\xi^+ \leftarrow \xi^+ + \nu^+$ would correspond to a nonempty localization set $\mathcal{X}$ that contains a ball of radius $r$. After this pre-processing step, we solved normally, as in the previous experiments, for an inner ellipsoid and an outer bounding box, using the updated noise bounds computed in the pre-processing step. Notice that such pre-processing step also serves as an outlier detection method, since whenever we find $\nu^- \neq 0$ or $\nu^+ \neq 0$ we know that some of the measurement errors did not respect the original bounds. The results of such experiments, averaged over the 100 runs, are reported in Table 7. These results are to be compared to those in Table 1, where we had the same assumed error bounds but no outliers.

| $m$ | $e(c_b)$ | $e(\hat{x})$ | $e_\%(c_b)$ | $e_\%(\hat{x})$ | $e_{\text{outer}}$ | $e_{\text{outer}\%}$ | $f_\%(c_b)$ | $f_\%(\hat{x})$ | $t_r$ | $\eta_{\text{vol}}$ |
|---|---|---|---|---|---|---|---|---|---|---|
| 3 | 84.18 | 74.70 | 9.78 | 8.42 | 30.60 | 3.70 | 54.55 | 80.81 | 1.58 | 2.07 |
| 4 | 65.94 | 59.64 | 8.06 | 7.34 | 15.64 | 1.94 | 50.00 | 66.33 | 1.58 | 5.91 |
| 5 | 43.16 | 42.27 | 5.37 | 5.26 | 11.47 | 1.39 | 52.00 | 69.00 | 1.71 | 5.41 |
| 6 | 20.07 | 20.19 | 2.52 | 2.53 | 6.87 | 0.89 | 53.61 | 70.10 | 1.83 | 7.79 |
| 7 | 17.48 | 19.99 | 2.17 | 2.48 | 5.91 | 0.74 | 44.90 | 59.18 | 1.96 | 10.49 |
| 8 | 15.83 | 15.76 | 1.92 | 1.91 | 4.74 | 0.60 | 52.17 | 57.61 | 2.08 | 13.62 |
| 9 | 14.21 | 14.17 | 1.83 | 1.82 | 3.52 | 0.44 | 48.91 | 56.52 | 2.19 | 15.95 |
| 10 | 13.27 | 13.74 | 1.72 | 1.79 | 3.17 | 0.40 | 61.29 | 64.52 | 2.27 | 16.83 |

TABLE 7
*Average results over $N = 100$ randomized tests, for $n = 2$, relative error level $\tilde{e}_i^+ = 0.2$ and 10% probability of outliers having 50% relative error.*

A few comments on the results of Table 7 are in order. First, we observe that the average running times $t_r$ are now slightly higher than those in Table 1, since they account for the additional effort due to the pre-processing, i.e., for solving Problem (6.6). Next, we observe that the presence of errors violating the assumed bounds tends to make $\mathcal{X}_{\text{true}}$ and $\mathcal{X}$ empty, while the pre-processing problem (6.6) aims at correcting this by enlarging the bounds so to make $\mathcal{X}$ nonempty. However, the bounds are adjusted so to make $\mathcal{X}$ "barely feasible," that is we seek for the minimum bound adjustment which guarantees that a *small* sphere is contained in $\mathcal{X}$. Therefore, even with pre-processing, in the presence of errors violating the assumed bounds we expect to have $\mathcal{X}$ sets smaller that those we obtained in Table 1 for the case where no outliers were present. Indeed, this fact is reflected in the columns $e_{\text{outer}}$ and $e_{\text{outer}\%}$ in Table 7 which measure the size of the bounding boxes, and appear to be smaller than the corresponding boxes in Table 1. The fact that the true feasible set $\mathcal{X}_{\text{true}}$ is smaller or empty in the presence of outliers is also reflected in the greater difficulty in finding a feasible point (cfr. columns $f_\%(c_b)$ and $f_\%(\hat{x})$) and in the higher volumetric ratios in column $\eta_{\text{vol}}$. The increased uncertainty due to the outlier errors naturally reflects also in higher estimation errors (absolute or relative distance from central estimate to the true target location), as reported in the first four columns of Table 7. We remark that in this latter numerical test we are using a model outside its intended scope, that is, under simulation conditions that severely violate the founding assumptions of the model itself (i.e., that the noise should be contained within the assumed bounds). Even in such "off-label" use, however, our proposed approach provides consistent results and permits validation, or invalidation, of the working assumptions on the noise bounds.

REMARK 8 (Comparison with point estimators). The literature offers various sophisticated point estimators for localization under Gaussian noise assumptions, such as the interior-point bias treatment in [16] or the diffusion Gauss-Newton method in [28]. While such methods effectively minimize metrics like the Root Mean Square Error (RMSE) of the point estimate, our approach fundamentally shifts the paradigm from point estimation to *guaranteed set outer bounding*. Under an interval unknown-but-bounded (UBB) noise model, a direct comparative analysis using RMSE is conceptually misaligned: our primary objective is to guarantee that the true location lies entirely within the computed set for *any* admissible noise realization, a property that stochastic point estimators are not designed to fulfill. Therefore, while point estimates can be extracted from our bounding sets (e.g., their geometric centers), comparing their RMSE against specialized Gaussian-noise point estimators falls outside the intended scope of the set-membership framework. ⋆

**8. Conclusions.** This paper discussed the problem of finding a set-valued estimate of an object's position from noisy but bounded range measurements. While standard location estimators only provide a single point estimate, the proposed approach returns a region (the localization set $\mathcal{X}$, or its outer bounding box or ellipsoid) which is *guaranteed* to contain the whole set $\mathcal{X}_{\text{true}}$ of the possible locations compatible with the measurements. On the one hand such guaranteed set-membership information can be useful and critical in certain applications, and on the other hand it has a clear geometrical meaning, and it can be computed efficiently and directly by means of convex programming, with no need to resort to linearizations or relaxations.

Several interesting aspects remain open to further investigation. For instance, the conservativeness of the inclusion $\mathcal{X}_{\text{true}} \subseteq \mathcal{X}$ could be explored formally, and its relation to the anchors' configuration geometry assessed quantitatively. Also, inner approximation of $\mathcal{X}_{\text{true}}$, or even just the determination of one point in this set, remains a problem worth additional investigation, although heuristics exists for its solution, e.g., linearization methods such as the one described in Section 6.1, or sampling-based methods. Further, one may consider the case where the anchors' locations are not known exactly, and investigate the effect of such uncertainty on the localization set.